\documentclass[preprint,showpacs,amsmath,amssymb,superscriptaddress]{revtex4}

\usepackage{graphicx}
\usepackage{mathrsfs}
\usepackage{dcolumn}
\usepackage{bm}

\newcommand{\NLTO}{Nd$_{2-x}$La$_{x}$Ti$_{2}$O$_{7}$}
\newcommand{\NTO}{Nd$_{2}$Ti$_{2}$O$_{7}$}
\newcommand{\RTO}{R$_{2}$Ti$_{2}$O$_{7}$}

\begin{document}

\title{Anisotropic paramagnetism of monoclinic Nd$_{2}$Ti$_{2}$O$_{7}$ single crystals}
\author{Hui Xing}
\affiliation{Department of Physics and State Key Laboratory of Silicon Materials, Zhejiang University, Hangzhou 310027, China}
\author{Gen Long}
\affiliation{Department of Physics, University at Buffalo, the State University of New York, Buffalo, New York 14260, USA}
\author{Hanjie Guo}
\affiliation{Department of Physics and State Key Laboratory of Silicon Materials, Zhejiang University, Hangzhou 310027, China}
\author{Youming Zou}
\affiliation{High Magnetic Field Laboratory, Chinese Academy of Sciences, Hefei 230031, China}
\author{Chunmu Feng}
\affiliation{Department of Physics and State Key Laboratory of Silicon Materials, Zhejiang University, Hangzhou 310027, China}
\author{Guanghan Cao}
\affiliation{Department of Physics and State Key Laboratory of Silicon Materials, Zhejiang University, Hangzhou 310027, China}
\author{Hao Zeng}
\email{haozeng@buffalo.edu}
\affiliation{Department of Physics, University at Buffalo, the State University of New York, Buffalo, New York 14260, USA}
\author{Zhu-An Xu}
\email{zhuan@zju.edu.cn}
\affiliation{Department of Physics and State Key Laboratory of Silicon Materials, Zhejiang University, Hangzhou 310027, China}

\begin{abstract}
The anisotropic paramagnetism and specific heat in \NTO\ single crystals are investigated. Angular dependence of the magnetization and Weiss temperatures show the dominant role of the crystal field effect in the magnetization. By incorporating the results from the diluted samples, contributions to Weiss temperature from exchange interactions and crystal field interactions are isolated. The exchange interactions are found to be ferromagnetic, while the crystal field contributes a large negative part to the Weiss temperature, along all three crystallographic directions. The specific heat under magnetic field reveals a two-level Schottky ground state scheme, due to the Zeeman splitting of the ground state doublet, and the $g$-factors are thus determined. These observations provide solid foundations for further investigations of \NTO.

\end{abstract}

\maketitle

\section{introduction}
Rare earth titanates (\RTO) are an interesting family of materials showing rich ferroelectric \cite{Shcherbakova1979,Winfield1992} and magnetic properties \cite{Gardner2010, Bramwell2009}. The common crystal structures \cite{Shcherbakova1979} for these materials are cubic pyrochlore for small rare earth ions (Sm$^{3+}$ \textendash\ Lu$^{3+}$) and monoclinic for large rare earth ions (La$^{3+}$ \textendash\ Nd$^{3+}$). The magnetic properties of the former have been explored extensively during the recent years \cite{Gardner2010}. The subtle interplay among the various interactions including crystal field, exchange and dipole-dipole, coupled with geometrical frustration specific to the pyrochlore lattice \cite{Moessner2006}, has led to exotic magnetic phases such as spin ice \cite{Harris1997, Ramirez1999, Bramwell2001} and spin liquid \cite{Gardner2001}. These phases are extremely sensitive to doping and external stimuli such as pressure \cite{Mirebeau2002} and magnetic fields \cite{Cao2008}. Interesting dynamic processes have also been identified \cite{Snyder2003, Xing2010}. On the other hand, the rare earth titanates with a monoclinic structure have received much less attention. Due to the removal of geometric frustration, these materials are believed to possess magnetically ordered phases at low temperatures. However, the competition between the crystal field and exchange interactions (plus dipole-dipole) with close energy scales may lead to interesting and unusual dynamic freezing processes towards the ordered states. In this regard, it is essential to retrieve the magnitude of the crystal field, exchange and the dipole-dipole interactions, in order to study the magnetic freezing processes.

Here we focus on the monoclinic \NTO\ (NTO). Some features of NTO, such as the ferroelectricity \cite{Winfield1992}, the optical magnetoelectric effect \cite{Shimada2008} and the photo-catalysis effect \cite{Hwang2003}, receive current interests in these materials. In this work, we study the dc susceptibility and specific heat of NTO single crystals. The anisotropic paramagnetism is investigated in detail. We found that crystal field interactions dominate the anisotropy, and exchange interactions $J_{ex}$ are estimated by combining the results of the diluted samples. The ground state of Nd$^{3+}$ spins splits into five Kramers doublets in the crystal field associated with the $C_{1}$ site in $P2_{1}(4)$ space group. The magnetic field splits the ground state doublet into two singlets, as is evidenced by the specific heat at different fields, showing a two-level Schottky scheme with the energy gap proportional to the field. These studies lay solid ground for further investigation of this system under doping and external perturbations.

\section{experiment}

Single crystals of \NTO\ and \NLTO\ were grown by the floating zone method in flowing oxygen. Seed rods for the single crystal growth were polycrystalline \NTO\ and \NLTO\ sintered by the standard solid state reaction in oxygen atmosphere, with starting materials of stoichiometric mixture of Nd$_{2}$O$_{3}$, La$_{2}$O$_{3}$ and TiO$_{2}$ powder (Alfa Aesar). Oxygen atmosphere was used to avoid oxygen deficiency. The lattice constants were obtained by fitting the peak positions of the x-Ray diffraction pattern on polycrystals obtained by crushing \NTO\ single crystals. The single crystals were cleaved along $a$ direction, and the in-plane $b$ and $c$ directions were determined by the Laue diffraction. Dc susceptibility were measured with a Quantum Design superconducting interference device magnetometer (SQUID MPMS-5) with fields of up to 5 T. Dc susceptibility at higher fields were measured by the vibrating sample magnetometer option in a Quantum Design physical property measurement system (PPMS). The in-plane angular dependence of the magnetization were measured using a commercial sample rotator in MPMS, the diamagnetic background of the sample holder was carefully subtracted. For the susceptibility measurement along different crystalline directions, the alignment of the crystal along the direction of the magnetic field was maintained within the error of $5^{\circ}$. Specific heat was measured in a PPMS using the thermal relaxation technique.

\begin{figure}
\includegraphics[width=0.95\columnwidth]{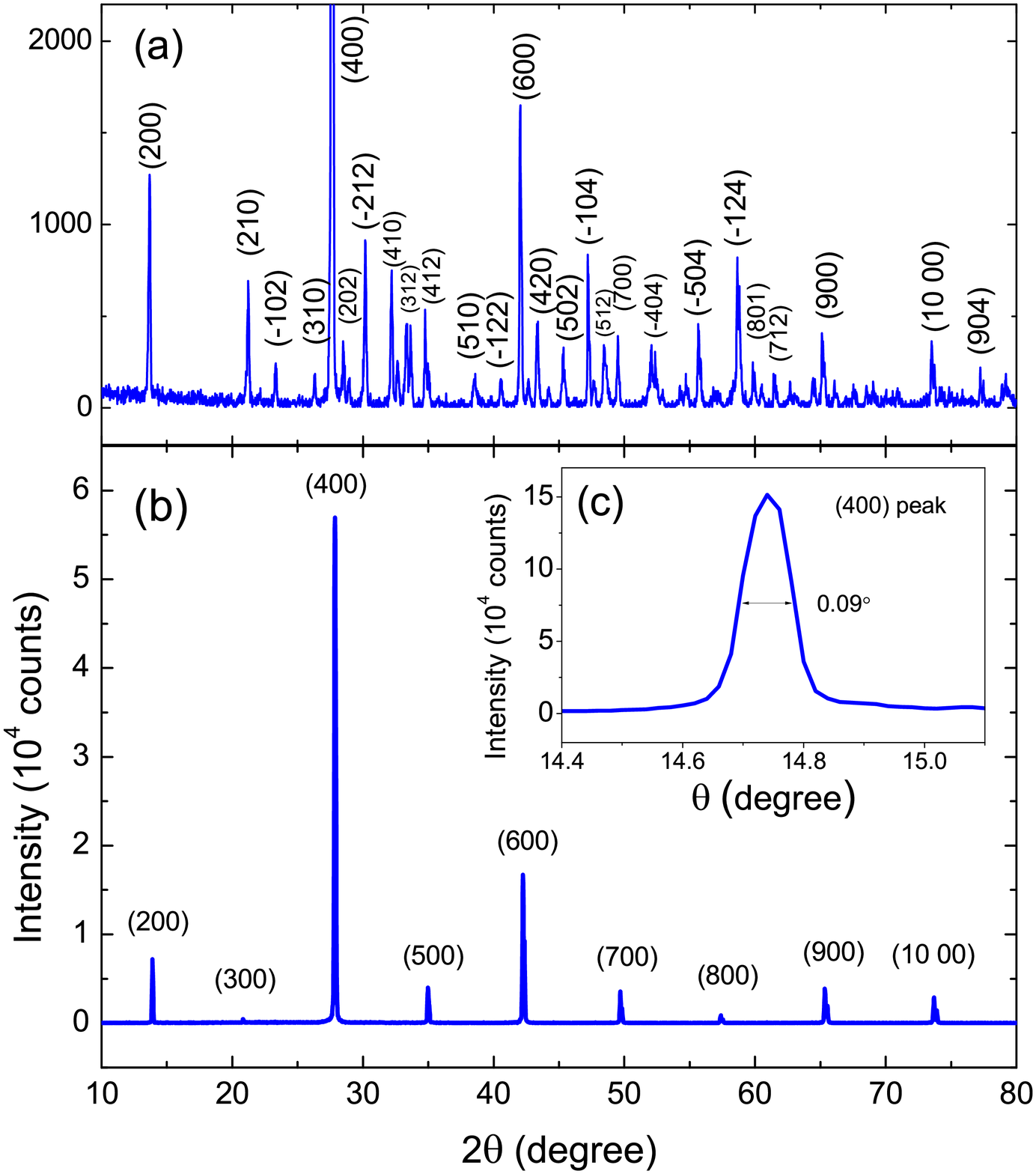}
\caption{\label{xrd}(color online)  (a) The x-ray diffraction pattern of the polycrystals obtained by crushing \NTO\ single crystals. Most peaks are indexed in the monoclinic structure in the space group $P2_{1}(4)$, the index of some small peaks are not marked explicitly.  (b) The x-ray diffraction pattern of a \NTO\ single crystal. Crystals cleave naturally  along the (100) direction. (c) The rocking curve of the (400) peak with FWHM = 0.09$^{\circ}$ indicates good crystallinity.}
\end{figure}

The as-grown NTO single crystals are of high quality and  large size benefiting from the floating zone method. The samples are monoclinic in space group $P2_{1}(4)$, the diffraction peaks of the polycrystals obtained by crushing NTO single crystals are well indexed as shown in Fig. \ref{xrd}(a), and resultant lattice parameters are $a = 13.002$ \AA , $b = 5.468$ \AA, $c = 7.678$ \AA, and $\alpha = \gamma = 90 ^{\circ}$, $\beta = 98.52 ^{\circ}$. Due to the layered structure, the crystal cleaves naturally along the (100) direction, as is shown in the XRD pattern in Fig. \ref{xrd} (b). In Fig. \ref{xrd} (c), the full width at half maximum (FWHM) of 0.09$^{\circ}$ of the rocking curve further indicates good crystallinity of the sample. The single crystals were cleaved and shaped into disks with typical dimensions of diameter $\sim$ 1.5 mm and thickness $\sim$ 0.2 mm. The disk shape mitigated the effect of demagnetization factor in the in-plane anisotropy of magnetization. However, we still corrected the demagnetization using the calculated value \cite{demag} $N = 0.75$ for $H \parallel a$ and 0.12 for $H \parallel bc$ respectively.

\section{results and discussions}
\subsection{Anisotropic Paramagnetism}

\begin{figure}
 \includegraphics[width=0.95\columnwidth]{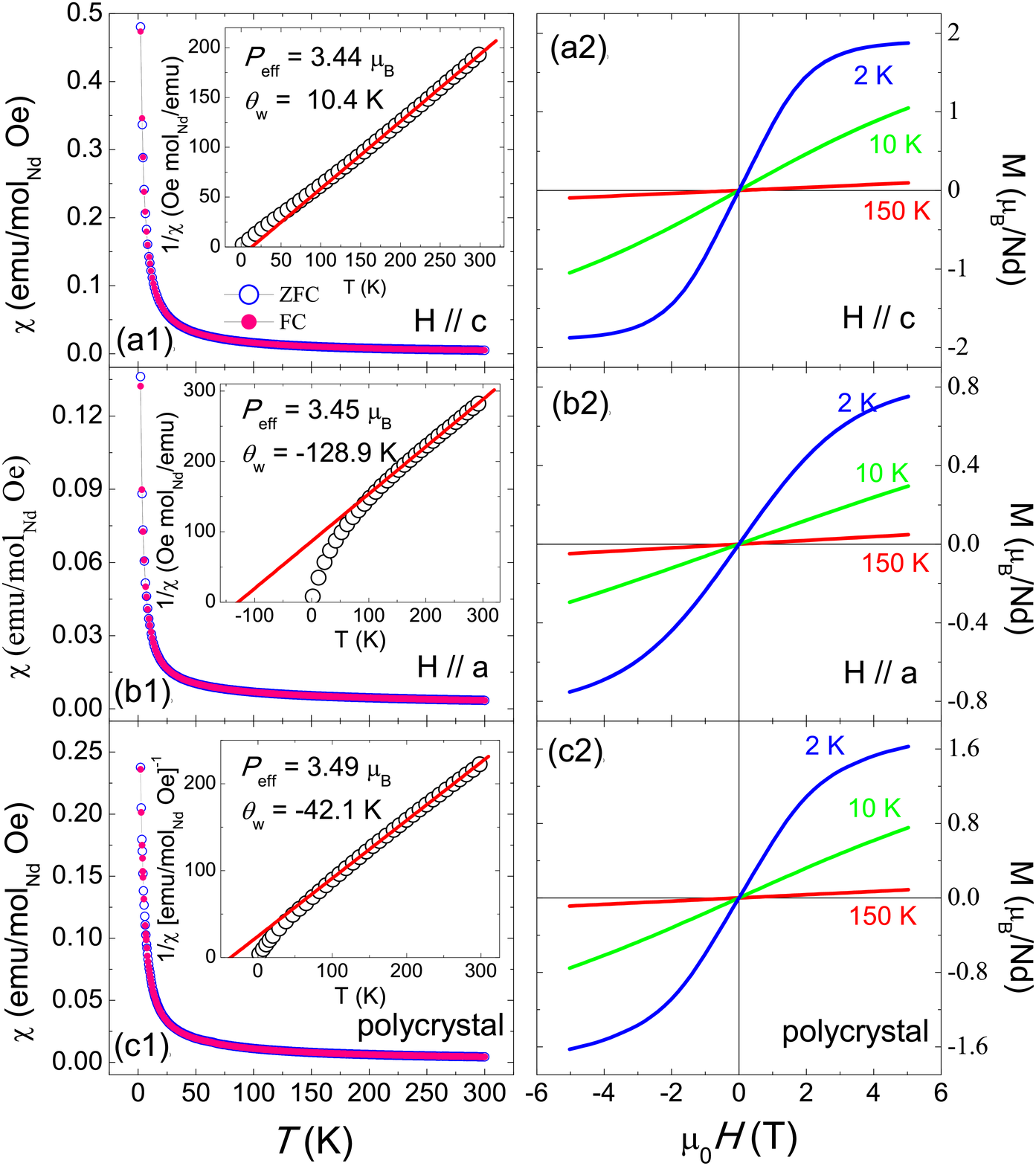}
\caption{\label{PM}(color online) (a1)-(c1): Zero-field-cooled and field-cooled susceptibility vs temperature at 1000 Oe for $H \parallel c$, $H \parallel a$ of a \NTO\ single crystal and a polycrystal sample, respectively. Insets are fittings by the the Curie-Weiss law.  (a2)-(c2): Isothermal magnetization curves at 2, 10, 150 K for $H \parallel c$, $H \parallel a$ and polycrystal respectively.}
\end{figure}

In Fig. \ref{PM} (a1)\textendash(c1), we show $\chi(T)$ at 1000 Oe with $H \parallel c$ and $H \parallel a$ for a single crystal and a polycrystal sample, respectively. The zero-field-cooled and field-cooled $\chi(T)$ overlap in the entire temperature range in study (2 \textendash\ 300 K), showing typical paramagnetic behavior. However, $\chi(T)$ is clearly anisotropic when comparing the behavior of the single crystal with  $H \parallel c$ and $H \parallel a$, as well as that of a polycrystal. Above 150 K, the system follows the Curie-Weiss law with the effective moment of 3.44 \textendash\ 3.49 $\mu_{B}$, close to the theoretical value of 3.62 $\mu_{B}$ (an experimental value \cite{Ashcroft} is 3.5 $\mu_{B}$). The resultant Weiss temperature $\theta_{W}$ varies from 10.4 K for $H \parallel c$ to $-128.9$ K for $H \parallel a$. $\theta_{W}$ for the polycrystal, which can be taken as the average effect from all the crystalline directions, is $-42.1$ K. The negative Weiss temperature of the polycrystal indicates an effective antiferromagnetic correlation between Nd$^{3+}$ spins. By comparing Fig. \ref{PM} (a1) with (b1), we see that the in-plane magnetization is more than three times larger than the $H \parallel a$ case, this suggests that the Nd$^{3+}$ spins prefer to align in the $bc$ plane. Fig. \ref{PM} (a2)\textendash(c2) show the isothermal magnetization curves for $H$ along different crystalline directions. As is expected for a paramagnetic system, $M(H)$ shows no hysteresis, is linear at high temperatures, and shows convex curvatures at low temperatures due to the spin polarization process. The $M(H)$ curves measured with the field applied along different directions again show strong anisotropy.

\begin{figure}
\centering \includegraphics[width=0.95\columnwidth]{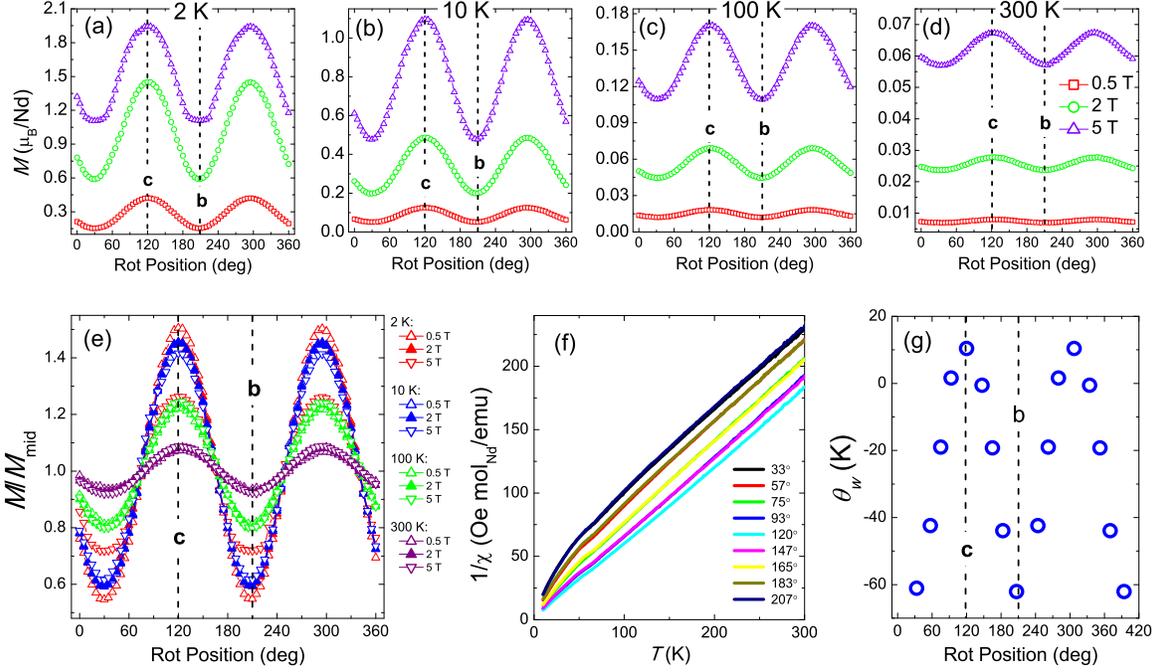}

\caption{\label{M-theta}(color online) (a)-(d): The angular-dependent in-plane magnetization $M(\theta)$ of \NTO\ at 2, 10, 100, 300K in the field of 0.5, 2, 5 T. (e): The angular dependence of the normalized magnetization $M/M_{mid}$, where $M_{mid}$ is the mean value of $M(\theta)$. (f) The temperature dependence of the inverse susceptibility of a \NTO\ single crystal with the in-plane field along $\theta =$ $33^{\circ}$, $57^{\circ}$, $75^{\circ}$, $93^{\circ}$, $120^{\circ}$, $147^{\circ}$, $165^{\circ}$, $183^{\circ}$ and $207^{\circ}$. (g): The in-plane angular dependence of the Weiss temperature by fitting $1/\chi(T)$ above 200 K in (f). Dashed lines labeled with $b$ and $c$ denote the in-plane crystalline $b$ and $c$ directions as determined by the Laue diffraction.}
\end{figure}

Moreover, the in-plane $M(H)$ curves themselves show anisotropy. Fig. \ref{M-theta} (a)\textendash(d) show the angular dependence of the in-plane magnetization. A two-fold symmetry is found at all fields and temperatures. The normalized in-plane anisotropy $M(\theta)/M_{mid}$, where  $M_{mid}$ is the mean value of $M(\theta)$, i.e., $\overline{\sum_{\theta} M(\theta)}$, is shown in Fig. \ref{M-theta} (e). The magnetization along the $c$ axis is roughly two times larger than that along the $b$ axis. As we already noted that the Nd$^{3+}$ spins prefer to align in the $bc$ plane, the $c$ axis is thus determined to be the easy axis of Nd$^{3+}$ spins.

We also show $1/\chi(T)$ with fields along different in-plane directions in Fig. \ref{M-theta} (f). $1/\chi(T)$ at different in-plane angles are parallel to each other at $T > 200$ K, as expected for paramagnetism with the same effective moments but different Weiss temperatures. The angular dependence of Weiss temperatures $\theta_{W}$ is shown in Fig. \ref{M-theta} (g).

Magnetic anisotropy, in particular magnetization anisotropy in paramagnets were previously reported in rare earth compounds \cite{Cho1996, Skumryev2009}. The primary origin of this anisotropy is the crystal electric field. When an ion is placed within a crystal lattice, the degeneracy of the Hund's rule ground state multiplet can be lifted by the electric field from the surrounding atoms. In a rare earth compound such as NTO, the cystal field can overcome the external field and pull the total moment resulting from the spin-orbit coupling into a certain direction. The anisotropic exchange coupling also contributes to the magnetic anisotropy. To separate the contribution of crystal field from exchange to the magnetic anisotropy, a series of diluted samples Nd$_{2-x}$La$_{x}$Ti$_{2}$O$_{7}$ were studied. Nonmagnetic La$^{3+}$ has very close ionic radius to that of Nd$^{3+}$, it dilutes the Nd$^{3+}$ spins and keeps the lattice almost unchanged. Thus, in a good approximation, the crystal field does not change with dilution while the exchange interaction decreases systematically with increasing La$^{3+}$ content. Because the Weiss temperature of the system reaches as low as $-150$ K for Nd$_{1.15}$La$_{0.85}$Ti$_{2}$O$_{7}$, and even lower for samples with lower Nd$^{3+}$ content, which makes it difficult to realize a reliable fitting using the Curie-Weiss law with the $M(T)$ data in the temperature range of 2 \textendash\ 300 K, only the data of Nd$_{1.15}$La$_{0.85}$Ti$_{2}$O$_{7}$ is presented here. The Nd$^{3+}$ content is determined indirectly by calculating the effective moment of the powder sample. By writing down the magnetic Hamiltonian, it is possible to extract both exchange and crystal field contributions to the angular dependent magnetization by assuming a constant crystal field contribution for the La$^{3+}$ diluted sample. Following the  approach in Ref. \cite{Boutron1973}, the magnetic hamiltonian consists of two parts, the exchange energy $\mathscr{H}_{ex}$ and the crystal field energy $\mathscr{H}_{CF}$ (the dipolar interaction can be assimilated in the exchange energy \cite{Boutron1973}). The exchange interaction is anisotropic and can be expressed as

\begin{equation}
\begin{array}{lcl}
\mathscr{H}_{ex} = -\sum_{\langle i, j\rangle} J_{ex}^{a} S_{i}^{a}S_{j}^{a}+J_{ex}^{b} S_{i}^{b}S_{j}^{b}+J_{ex}^{c} S_{i}^{c}S_{j}^{c}
\end{array}
\label{eq-exchange}
\end{equation}
where $a, b, c$ denote the three crystalline directions, and $J_{ex}^{a}$, $J_{ex}^{b}$ and $J_{ex}^{c}$ are the exchange constants for Nd$^{3+}$ spins along $a, b$ and $c$ directions, respectively. The crystal field part, for the low-symmetry $C_{1}$ site of the Nd$^{3+}$, is known to be
\begin{equation}
\begin{array}{lcl}
\mathscr{H}_{CF} = \sum_{p = 2, 4, 6} \sum_{k = -p}^{p} B_{p}^{k} O_{p}^{k}
\end{array}
\label{eq-cf}
\end{equation}
where $B_{p}^{k}$ are crystal field parameters, and $O_{p}^{k}$ are Stevens equivalent operators. It is known that the high temperature susceptibility of a system with only one spin species can be expressed as \cite{Boutron1973}

\begin{equation}
\begin{array}{rcl}
\frac{1}{\chi} = \frac{1}{C}(T+c\frac{2J(J+1)}{3Nk}\frac{\overrightarrow{H}\cdot \overleftrightarrow{N}\cdot \overrightarrow{H}}{H^{2}}+\frac{(2J-1)(2J+3)}{10k}\frac{\overrightarrow{H}\cdot \overleftrightarrow{u}\cdot \overrightarrow{H}}{H^{2}})\\+O(\frac{1}{T}).
\end{array}
\label{eq-ki-1}
\end{equation}
where $C$ is the Curie constant, $c$ is the percentage of the magnetic ions on the spin lattice, in Nd$_{2-x}$La$_{x}$Ti$_{2}$O$_{7}$ case, $c = (2-x)/2$. $\overleftrightarrow{N}$ and $\overleftrightarrow{u}$ are the exchange tensor and the crystal field tensor, and can be expressed by the parameters in eq. \ref{eq-exchange} and eq. \ref{eq-cf} respectively. When the three axes of the crystal are of at least second-order symmetry (symmetry no lower than $L_{2}$), eq. \ref{eq-ki-1} can be expressed with the parameters in eq. \ref{eq-exchange} and \ref{eq-cf} along each crystalline direction as

\begin{equation}
\begin{array}{lcl}
\frac{1}{\chi_{\alpha}} = \frac{1}{C}(T-c\frac{J(J+1)}{3} J_{ex}^{\alpha}+\frac{(2J-1)(2J+3)}{10k}u_{\alpha\alpha})
\end{array}
\label{eq-ki}
\end{equation}
here, $\alpha$ denotes the crystalline direction $a$, $b$ or $c$, $J_{ex}^{\alpha}$ are the exchange constants as already appeared in eq. \ref{eq-exchange}, and $u_{\alpha\alpha}$ is the $\alpha$ direction component of tensor $\overleftrightarrow{u}$. In specific symmetries, $u_{\alpha\alpha}$ can be expressed explicitly with the crystal field parameters $B_{p}^{k}$. The Weiss temperature is then

\begin{equation}
\begin{array}{lcl}
\theta_{W}^{\alpha} = c\frac{J(J+1)}{3} J_{ex}^{\alpha}-\frac{(2J-1)(2J+3)}{10k}u_{\alpha\alpha}
\end{array}
\label{eq-theta}
\end{equation}
However, in the NTO case, the symmetry of Nd$^{3+}$ site is barely $C_{1}$, an explicit expansion of $\overleftrightarrow{u}$ is thus difficult. Therefore only the exchange parameter $J_{ex}$ is extracted and the second term in Eq. \ref{eq-theta} is left as a single parameter $\theta_{CF}$ to be determined.

 The results of the Curie-Weiss fitting for diluted samples suggest that the contributions to the Weiss temperature from the exchange interactions $\theta_{ex}^{\alpha} = c\frac{J(J+1)}{3} J_{ex}^{\alpha}$ and crystal field interactions $\theta_{CF}^{\alpha} = -\frac{(2J-1)(2J+3)}{10k}u_{\alpha\alpha}$
are of opposite signs. The fitted Weiss temperature $\theta_{W}$, and the contributions from exchange interaction and crystal field $\theta_{ex}$ and $\theta_{CF}$ in all three crystallographic orientations for NTO and Nd$_{1.15}$La$_{0.85}$Ti$_{2}$O$_{7}$ are listed in Table I. With these Weiss temperatures, the exchange constants are determined to be $J_{ex}^{a}= 7.1\pm 1.1$ K, $J_{ex}^{b}= 6.9 \pm 1.1$ K and $J_{ex}^{c}= 4.2 \pm 0.7$ K. It can be seen that the exchange constants are positive in all crystallographic orientations. This illustrates the dominant role of the crystal field contribution to the magnetic anisotropy. While the exchange interactions are always ferromagnetic, the apparent exchange deduced from the Weiss temperature in different directions can be either positive or negative. This is because the crystal field contributes a large negative part to the Weiss temperature. As seen from Table I, $\theta_{CF}$ is a couple of times larger in magnitude than $\theta_{ex}$ along $a$ and $b$ axes and comparable in magnitude along the $c$ axis, making the Weiss temperature large and negative along $a$, $b$ directions and small but positive along the $c$ direction.

 It is also worth to note that eq. \ref{eq-ki} is independent of the magnetic field. Indeed, we see that in Fig. \ref{M-theta} (e), the normalized in-plane anisotropy $M(\theta)/M_{mid}$ is nearly constant for different fields at the same temperature. Only when spins are polarized significantly  by the field, i. e., M(H) curve is not linear anymore, $M(\theta)/M_{mid}$ at different fields bifurcate, as is the case at 2 K.

 As we have discussed, the monoclinic NTO does not accommodate geometrical frustration. However, with the absence of a magnetic transition down to at least 2 K, and the average Weiss temperature $\theta_{W} \sim -42.1$ K, the frustration index as defined by $f = |\theta_{W}|/T_{c}$ (Ref. \cite{Ramirez1994}) of this system reaches about 21, and is even higher along $a$ and $b$ directions. The large frustration index may indicate an unusual ground state and interesting spin freezing processes. These investigations are currently under way.

\subsection{Specific Heat}

\begin{figure}
\centering \includegraphics[width=0.95\columnwidth]{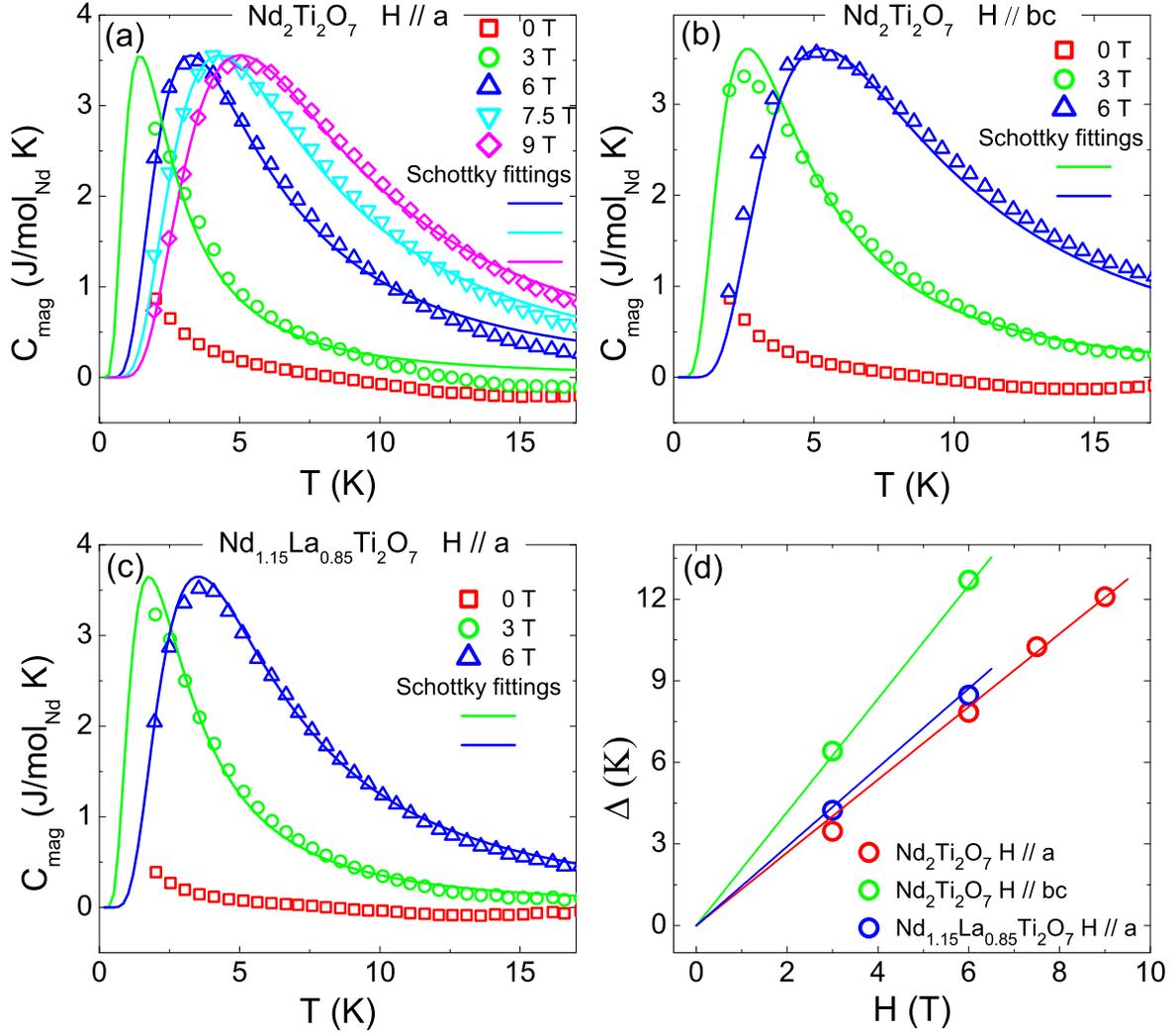}

\caption{\label{specific-heat}(color online) Magnetic specific heat of (a) A \NTO\ single crystal with $H \parallel a$; (b) A \NTO\ single crystal with $H \parallel bc$ (the field is not applied intentionally along a specific in-plane direction); (c) Nd$_{1.15}$La$_{0.85}$Ti$_{2}$O$_{7}$ with $H \parallel a$. Solid lines in (a)-(c) are two-level Schottky fittings; (d) The energy gap $\Delta$ obtained by the two-level Schottky fittings.  }
\end{figure}

The magnetic ground state of $^{4}I_{9/2}$ Nd$^{3+}$ on the $C_{1}$ site in the $P2_{1}(4)$ space group should split into five Kramers doublets due to the crystal field. Calculations on the Nd$^{3+}$ in the same symmetry in other compounds indeed revealed the five doublets scheme \cite{Agnieszka2008}, while a detailed calculation on NTO is still lacking. The specific heat measurement provides more information on the magnetic ground state. Fig. \ref{specific-heat} (a)\textendash(c) show the magnetic specific heat of NTO for H $\parallel$ a, H $\parallel$ bc and that of Nd$_{1.15}$La$_{0.85}$Ti$_{2}$O$_{7}$ for H $\parallel$ a at different fields. The magnetic specific heat is obtained by subtracting the lattice contribution using the specific heat data of a nonmagnetic La$_{2}$Ti$_{2}$O$_{7}$ single crystal. The zero-field specific heat in principle detects any magnetic phase transitions, and the magnetic entropy associates directly with the ground state degeneracy. In Fig. \ref{specific-heat} (a), the zero-field magnetic specific heat shows no anomaly for a magnetic transition, which is consistent with the paramagnetic behavior above 2 K. There is a progressive upturn as $T$ decreases, which signals the presence of a large portion of the magnetic entropy below 2 K. The magnetic entropy is therefore difficult to integrate due to the limited temperature range. However, this problem is bypassed by using the specific heat in a magnetic field. As shown in Fig. \ref{specific-heat} (a), $C_{mag}(T)$ in a field shows a broad peak with the magnitude of around 3.6 J/mol$\cdot$ K at low temperatures, and the maximum moves to higher temperatures with its magnitude almost unchanged as the field increases. These indicate that the peaks are Schottky anomalies associated with the ground state doublet splitting in the presence of a magnetic field. The two-level Schottky specific heat is,
\begin{equation}
\begin{array}{lcl}
C_{Schottky} = R (\frac{\Delta}{T})^{2} \frac{e^{\Delta/T}}{(1 + e^{\Delta/T})^{2}}
\end{array}
\label{eq3}
\end{equation}
where $R$ is the gas constant, and $\Delta$ is the energy gap between the two levels.

Eq. \ref{eq3} describes the data nicely as shown by the solid lines in Fig. \ref{specific-heat} (a). It can be seen in Fig. \ref{specific-heat} (d) that the fitting parameter $\Delta$ scales linearly with magnetic field. This further proves that the gap between the two Schottky levels are caused by the zeeman splitting of the ground state doublet. Using $\Delta = g \mu_{B} H$, the $g$ factor for NTO with $H$ along $a$ is found to be 1.99. The magnetic specific heat is also studied in \NTO\ with $H \parallel bc$ and in Nd$_{1.15}$La$_{0.85}$Ti$_{2}$O$_{7}$ with $H \parallel a$. For NTO with $H \parallel bc$, the $g$ factor is 3.10. The anisotropy in $g$ factor again points to  the presence of strong crystal field, and these $g$ values should be corroborated with future crystal field calculations.

\section{conclusion}

 In conclusion, we have studied both single- and poly-crystalline \NTO\ with angle-dependent dc susceptibility and specific heat. The anisotropic paramagnetism is observed and is mainly attributed to the crystal field effect. The easy axis is determined to be the $c$-axis in the $bc$ plane. The contributions to Weiss temperature from exchange interactions and crystal field interactions are separated.  The exchange interactions are ferromagnetic, while the crystal field contributes a large negative part to the Weiss temperature, along all three crystallographic directions. This leads to large negative $\theta_{W}$ values along $a$ and $b$ axes and small positive $\theta_{W}$ along the $c$ axis.  The $^{4}I_{9/2}$ ground state multiplet of Nd$^{3+}$ splits into five Kramers doublets. The ground state doublet further splits into two singlets in a magnetic field, and this two-level ground state scheme is revealed by the magnetic specific heat. $g$ factors show anisotropy too, again due to the crystal field. These studies provide solid foundations for further investigations of this system, including the magnetic ground state and the spin freezing processes.

\begin{acknowledgments}
 This work is supported by the NSFC (Grant No. 10634030), US NSF DMR-0547036, the Fundamental Research Funds for the Central
Universities of China (Grant No. 2010QNA3026), and National Basic
Research Program of China (Grant No. 2007CB925001).
\end{acknowledgments}

\pagebreak

\begin{table}[tbph]

\begin{center}
\begin{tabular}{|c|c|c|c||c|c|c||c|c|c|}
\hline
    Unit: K & $\  \ \ \theta_{W}^{a}$ \ \ \  & $\ \ \ \theta_{ex}^{a}$ \ \ \ & $\ \ \ \theta_{CF}^{a}$ \ \ \ &\ \ \  $\theta_{W}^{b}$ \ \ \ & $\ \ \ \theta_{ex}^{b}$ \ \ \  &\ \ \  $\theta_{CF}^{b}$ \ \ \ & \ \ \ $\theta_{W}^{c}$ \ \ \ &\ \ \  $\theta_{ex}^{c}$ \ \ \ &\ \ \  $\theta_{CF}^{c}$ \ \ \  \\ \hline
Nd$_{2}$Ti$_{2}$O$_{7}$ & $-128.9$ & $58.3$ & $-187.5$ & $-62.1$ & $57.0$ & $-119.1$ & $10.4$ & $34.3$ & $-23.8$ \\ \hline
Nd$_{1.15}$La$_{0.85}$Ti$_{2}$O$_{7}$ & $-153.8$ & $33.7$ & $-187.5$ & $-86.3$ & $32.8$ & $-119.1$ & $-4.1$ & $19.7$ & $-23.8$ \\ \hline

\end{tabular}
\caption{The Weiss temperatures $\theta_{W}$ for Nd$_{2}$Ti$_{2}$O$_{7}$ and Nd$_{1.15}$La$_{0.85}$Ti$_{2}$O$_{7}$ along $a$, $b$ and $c$ directions. The fitting errors are estimated to be within 3 K. The contributions from the exchange interactions and crystal field interactions are isolated as $\theta_{ex}$ and $\theta_{CF}$ using eq. \ref{eq-theta}. }
\end{center}
\end{table}

\end{document}